\documentclass{article}

\usepackage{arxiv}

\usepackage[utf8]{inputenc}
\usepackage[T1]{fontenc}    
\usepackage{hyperref}       
\usepackage{url}            
\usepackage{booktabs}       
\usepackage{amsfonts}       
\usepackage{amsmath}
\usepackage{nicefrac}       
\usepackage{microtype}                   
\usepackage{graphicx}
\usepackage{graphicx}
\usepackage{verbatim}

\usepackage{amsmath,amssymb,amsthm}
\newtheorem{proposition}{Proposition}

\usepackage{tikz}
\usetikzlibrary{calc}

\usepackage{multirow}

\usepackage{enumitem}
\setlist[itemize]{labelindent=1em,leftmargin=2em,labelsep=0.5em}

\usepackage{braket}

\makeatletter
\def\blfootnote{\xdef\@thefnmark{}\@footnotetext}
\makeatother

\usepackage{fancyhdr}
\begin{document}

\title{Encrypted clones can leak: Classification of informative subsets in Quantum Encrypted Cloning}

\author{
  \href{https://orcid.org/0000-0001-5186-0199}{\includegraphics[scale=0.06]{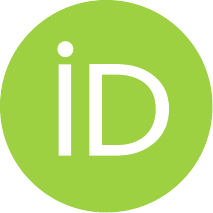}\hspace{1mm} Gabriele Gianini}\\
  Universit\`a degli Studi di Milano-Bicocca \\
  Milan, Italy \\
  \texttt{gabriele.gianini@unimib.it}\\
\and
\href{https://orcid.org/0000-0002-9585-7810}{\includegraphics[scale=0.06]{orcid.pdf}\hspace{1mm}
  Omar Hasan}\\
  Institut National des Sciences Appliqu\'ees de Lyon\\
  Lyon, France \\
  \texttt{omar.hasan@liris.cnrs.fr}\\ \vspace{1pt}
\and
  \href{https://orcid.org/0000-0002-1087-4866}{\includegraphics[scale=0.06]{orcid.pdf}\hspace{1mm}
  Corrado Mio}\\
  Khalifa University of Science and Technology\\
  Abu Dhabi, UAE\\
\texttt{corrado.mio@ku.ac.ae}\\
  \and
  \href{https://orcid.org/0000-0003-1737-6218}{\includegraphics[scale=0.06]{orcid.pdf}\hspace{1mm}
  Stelvio Cimato}\\
  Universit\`a egli Studi di Milano\\
  Milan, Italy\\
  \texttt{stelvio.cimato@unimi.it}\\ \vspace{1pt}
    \and
  \href{https://orcid.org/0000-0002-9557-6496}{\includegraphics[scale=0.06]{orcid.pdf}\hspace{1mm}
  Ernesto Damiani}\\
  Universit\`a degli Studi di Milano\\
  Milan, Italy\\
  \texttt{ernesto.damiani@unimi.it}
}
\date{}
\maketitle
\begin{abstract}
Encrypted cloning enables the redundant storage of an unknown qubit while remaining compatible with the no-cloning theorem, since only one clone can later be recovered through key-consuming decryption. Because encryption in this protocol is introduced to enable cloning-compatible redundancy rather than to guarantee confidentiality by design, its secrecy properties must be assessed explicitly. Here we classify the subsets of the encrypted-clone storage register into authorized, completely non-informative, and partially informative sets. We show that intermediate non-authorized subsets may retain only a restricted residual dependence on the input state, and we characterize exactly when this dependence occurs. The resulting leakage pattern is parity-dependent, revealing a structural confidentiality limitation of encrypted cloning. 
\keywords{
Encrypted cloning;
Quantum redundancy;
Information leakage;
Authorized subsets;
Confidentiality
}
\end{abstract}


\section{Introduction}\label{sec:intro}

The no-cloning theorem is a fundamental obstacle to quantum storage, as it forbids the direct replication strategies that support reliability in classical systems. Recent work by Yamaguchi and Kempf \cite{yamaguchi2026encrypted,yamaguchi2026experimental}, however, introduced \textit{encrypted cloning}, a Pauli-operator-based protocol that enables redundant storage of an unknown quantum state while remaining compatible with the no-cloning theorem: although multiple encrypted clones are created, only one can later be recovered through key-consuming decryption.

Encrypted cloning thus introduces a new primitive for quantum information processing: \textit{redundant availability without readable duplication}. This is especially attractive for storage, where redundancy supports resilience and availability, yet direct copying is forbidden. Starting from a source qubit \(A\) and \(n>1\) Bell pairs, the protocol applies a Pauli-based unitary transformation and produces a storage register composed of \(n\) clone qubits and \(n\) noise qubits.

A central feature of the protocol is that each clone qubit, taken in isolation, is non-informative. Perfect recovery is possible only from authorized subsets of the storage register, namely those containing at least \(n+1\) qubits, including one complete clone-noise pair and at least one qubit from every remaining pair. In this respect, encrypted cloning resembles access-structured protocols such as quantum secret sharing: recoverability is governed not by the mere presence of multiple clones, but by the structure of the subset that is accessed.

Yet recoverability and confidentiality are not the same issue. In encrypted cloning, encryption is introduced primarily as the mechanism that makes redundancy compatible with quantum constraints, not as a confidentiality primitive optimized in its own right. For storage applications, it is therefore not enough to know that isolated clones reveal nothing and that authorized subsets allow perfect recovery; one must also determine what can be learned from intermediate, non-authorized subsets of clone and noise qubits. If such subsets retain partial information about the input state, then encrypted cloning does not provide an all-or-nothing confidentiality guarantee.

In this work, we provide a complete classification of subsets of the encrypted-clone storage register into \textit{authorized}, \textit{completely non-informative}, and \textit{partially informative} sets. We show that intermediate non-authorized subsets are not all equivalent: while some reveal no information about the input state, others retain a restricted residual dependence on it, occurring only through its \(y\)-Bloch component. The resulting leakage pattern is parity-dependent, revealing a structural confidentiality limitation of encrypted cloning.

The remainder of the paper is organized as follows. Section~2 recalls the encrypted-cloning protocol and introduces the criterion used to distinguish authorized, completely non-informative, and partially informative subsets. Section~3 analyzes the low-dimensional cases \(n=2\) and \(n=3\), where the leakage mechanism becomes explicit. Section~4 provides the general classification of subsets of the encrypted-clone storage register. Section~5 concludes with a discussion of the confidentiality implications of these results.

\section{Formal background and problem setup}

We recall the encrypted-cloning protocol introduced in \cite{yamaguchi2026encrypted}, restricting attention to the case of a single input qubit and to the notation needed for leakage analysis.

\subsection{The Pauli-based encoding}
Let \(A\) denote the input qubit, prepared in an unknown pure state $\ket{\psi}_A$.
To generate \(n>1\) encrypted clones, the protocol introduces \(n\) Bell pairs
\[
\ket{\phi}_{S_iN_i}=\frac{1}{\sqrt 2}(\ket{00}+\ket{11}),
\qquad i=1,\dots,n,
\]
where \(S_i\) is the \(i\)-th signal qubit and \(N_i\) is the corresponding noise, or key, qubit. 

The encrypted-cloning encoding acts jointly on the input qubit \(A\) and on the signal qubits \(S_1,\dots,S_n\), while leaving the noise qubits \(N_1,\dots,N_n\) unaffected. To encode a pure state $\ket{\psi}$ into $n$ encrypted clones, Yamaguchi and Kempf use the following unitary,  
based on the Pauli operators:
\(
\sigma_0=I,\,
\sigma_1=X,\,
\sigma_2=Y,\,
\sigma_3=Z.
\)
\[
U_{\mathrm{enc}}^{(n)}
=
\frac{1}{2}\sum_{\mu=0}^3 \alpha_\mu^{-1}\sigma_\mu^{(A)}
\otimes\left(\bigotimes_{i=1}^n \sigma_\mu^{(S_i)}\right),
\]
with
\begin{equation}\label{eq:alphas}
\alpha_0=1,\qquad
\alpha_1=\alpha_3=i,\qquad
\alpha_2=-i^{\,n+1}.
\end{equation}

The resulting encoded state (see \cite{yamaguchi2026encrypted}, Eq.~(7)) is
\[
\ket{\Psi_{\mathrm{enc}}}
=
U_{\mathrm{enc}}^{(n)}
\left[
\ket{\psi}_A \otimes
\left(\bigotimes_{i=1}^n \ket{\phi}_{S_iN_i}\right)
\right]
=
\frac{1}{2}\sum_{\mu=0}^3 \alpha_\mu^{-1}\sigma_\mu^{(A)}\ket{\psi}_A
\otimes
\left(\bigotimes_{i=1}^n \ket{\phi_\mu}_{S_iN_i}\right),
\]
where we use the shorthand
\[
\ket{\phi_\mu}_{S_iN_i}
\equiv
\left(\sigma_\mu^{(S_i)}\otimes I^{(N_i)}\right)\ket{\phi}_{S_iN_i}.
\]

The corresponding density matrix is
\begin{equation}\label{eq:encoded-state}
\rho_{\mathrm{enc}}^{(n)}
=
\frac{1}{4}
\sum_{\mu,\nu=0}^3
\alpha_\mu^{-1}\alpha_\nu\,
\left(\sigma_\mu^{(A)} \ket{\psi}\!\bra{\psi}\,\sigma_\nu^{(A)}\right)
\otimes
\left(
\bigotimes_{i=1}^n
\ket{\phi_\mu}\!\bra{\phi_\nu}_{\,S_iN_i}
\right).
\end{equation}

This decomposition shows that encrypted cloning is governed by the coherent superposition of four Pauli branches, which gives rise to sixteen terms in the density-matrix expansion. Yamaguchi and Kempf formulate the protocol for \(n>1\). The case \(n=1\) is considered exceptional, since the single clone \(S_1\) is not fully encrypted; we return to this point in Section~3.

\subsection{Authorization vs. informativeness}

Yamaguchi and Kempf \cite{yamaguchi2026encrypted} showed which sets are authorized and which are unauthorized. Authorized sets are fully informative, i.e. hold enough information to enable the full recovery of the original state.

\paragraph{Authorization.} The authorized sets are defined by the following rules: 
\begin{itemize}
\item[(AUTH1)] A subset is authorized if it contains one complete signal-noise pair and at least one qubit from each of the remaining \(n-1\) pairs. For instance,
\begin{itemize}
\item
the set containing one encrypted clone qubit and all the noise qubits is authorized;
\end{itemize}
\item[(AUTH2)] 
Equivalently, a subset is unauthorized whenever it fails to satisfy this condition. For instance,
\begin{itemize}
\item
any subsystem consisting of \((n-1)\) complete pairs is unauthorized
\end{itemize}
\item[(AUTH3)]
Any subsystem consisting of all \(n\) noise qubits is unauthorized.
\end{itemize}

\paragraph{Informativeness.} Crucially, however, unauthorized access (i.e., inability to recover the original qubit) does not by itself imply complete absence of leakage nor maximal mixedness. In principle, sub-authorized subsets may still retain partial information about the input state. 

The original work already identified some unauthorized subsets that are completely uninformative: 
\begin{itemize}
\item[(INF1)]
for $n>1$ each individual clone is maximally mixed, and therefore locally uninformative;
\item[(INF2)]
the source qubit \(A\) is likewise maximally mixed after encoding.
\end{itemize}

What remains to be determined is whether other unauthorized subsets are fully independent of the input state or instead exhibit residual leakage.
Whether a reduced subsystem is fully non-informative or instead retains partial information about the input state is determined by the interference structure among the Pauli branches in equation (\ref{eq:encoded-state}).

\subsection{Statement of the problem and scope of the analysis}

Given the set of $2n$ qubits making up the \textit{storage register} defined by 
$$
{\mathcal R}_n\equiv \{S_1,N_1,S_2,N_2,\dots,S_n,N_n\}
$$
Consider the subsets $B \subseteq {\mathcal R}_{n}$.

\subsubsection{Statement of the problem}
 The work by Yamaguchi and Kempf \cite{yamaguchi2026encrypted} tells which subsets are \textit{fully authorized}. Our aim is to determine which are \textit{completely uninformative}, i.e. independent of the input state \(\ket{\psi}\), and which are instead \textit{partially informative}.
For the partially informative subsets, we also want to provide an expression of the reduced state, to characterize the leakage.\paragraph{Leakage criterion.}
Given a subsystem \(B\), we call it \textit{completely uninformative} if its reduced state
\(
\rho_B(\psi)
\)
is independent of the input state \(\ket{\psi}\). In contrast, we say that \(B\) exhibits leakage if \(\rho_B(\psi)\) retains a nontrivial dependence on \(\ket{\psi}\), even when \(B\) is not an authorized subset and does not allow full recovery.

A special case of \textit{completely uninformative} state is a maximally mixed state. However, we will meet completely uninformative states which are not maximally mixed.

\subsection{Subsets missing a full pair}

A  class of subsets important for our discussion are those that miss a complete signal-noise pair $\{S_i,N_i\}$. Yamaguchi and Kempf just state that they are unauthorized, however it is easy to show that they are also \textit{completely uninformative}. This will help delimit the scope of our analysis of the reduced states.

\begin{proposition}
Any subset \(B\subset \mathcal R_n\) that misses a complete pair \(\{S_j,N_j\}\) is completely non-informative.
\end{proposition}

\paragraph{Tracing off $A$.} The full encoded state $\rho_{\mathrm{enc}}^{(n)}$ consists of the qubit $A$ (no longer in the input state $\ket{\psi}$, since transformed by the encoding) and of $n$ pairs $\{S_i,N_i\}$. To obtain the reduced state of any subset of ${\mathcal R}_n$ we will have to trace-off $A$ from expression (\ref{eq:encoded-state}). Thus the factor corresponding to $A$ will be replaced by
\[
\operatorname{Tr}_A\left(
\sigma_{\mu} \ket{\psi}\bra{\psi}\sigma_{\nu}\right)
=
\operatorname{Tr}_A\left(
\bra{\psi}\sigma_{\nu}\sigma_{\mu} \ket{\psi}\right)
=
\bra{\psi}\sigma_{\nu}\sigma_{\mu} \ket{\psi}
\]
It is worth noticing that for $\mu=\nu$ this scalar is identically equal to $1$ for the property of the Pauli operators and the normalization of the state $\ket{\psi}$.

Notice as well that for $\mu=\nu$ also $\alpha_{\nu}^{-1}\alpha_{\mu}=1.$

\paragraph{Tracing out a full pair $\{S_j,N_j\}$ from the storage register.}
Removing a full pair $\{S_j,N_j\}$ from ${\mathcal R}_n$ implies tracing-off both $S_j$ and $N_j$ from the factor $\ket{\phi_\mu}\bra{\phi_\nu}_{\{S_jN_j\}
}$ in expression (\ref{eq:encoded-state}).
Thus,
\[
\operatorname{Tr}_{S_j N_j}\left(\left|\phi_\mu\right\rangle\left\langle\phi_\nu\right|\right)=\left\langle\phi_\nu \mid \phi_\mu\right\rangle =\delta_{\mu \nu},
\]
due to the orthonormality of the Bell basis. This removes from (\ref{eq:encoded-state}) all the terms with $\mu\neq \nu$. If we have traced-off also $A$, then, thanks for the fact that the coefficients are equal to $1$, we are left with
$$
\rho_{\mathrm{red}}=\frac{1}{4} 
\sum_{\mu=0}^3 
\bigotimes_{i \neq j}\,
\Big(\left|\phi_\mu\right\rangle
\left\langle\phi_\mu\right|\Big)_{S_i N_i}
$$

This state does not carry any dependence on the initial state $\ket{\psi}$, therefore it is completely uninformative about it.

In other words, the encoded state (\ref{eq:encoded-state}) was carrying a dependence from $A$ through the factor $\big(
\sigma_\mu|\psi\rangle\langle\psi| \sigma_\nu
\big)$. If we trace-out only $A$ this factor is transformed into $\bra{\psi}\sigma_{\nu}\sigma_{\mu} \ket{\psi}$, which still preserves a dependence through the off diagonal elements (since the diagonal ones are equal to $1$);\ however, if we trace out a full pair, the corresponding factors are replaced by $\delta_{\mu\nu}$, which wipes away the off diagonal terms, leaving no dependence on $\ket{\psi}$ in the residual state. Although the reduced state is not maximally mixed, it is \textit{completely uninformative} about $\ket{\psi}$.

Now, if we remove further qubits from the set, i.e. we trace-off further qubits, no dependence on $\ket{\psi}$ is recovered. Thus, any subset of $B\subset{\mathcal R}_n$ missing even a single full pair  $\{S_j,N_j\}$ is completely uninformative about the initial state $\ket{\psi}$.
In summary, we have another criterion of informativeness
\begin{itemize}
\item[(INF3)] Subsets of the storage register missing a complete pair $\{S_j,N_j\}$ are completely uninformative about $\ket{\psi}$.
\end{itemize}

\subsection{Scope of the analysis}

\paragraph{The storage register subsets with $|B|> n$.} We know that, by construction, the full storage register ${\mathcal R}_n$ is authorized, and that if we remove one by one the qubits from it, caring that at least one qubit from each pair is present and at least a full pair is present, we still get an authorized set.
Thus, we do not have to investigate the subsets $B\subset {\mathcal R}_n$ with $|B|\geq (n+1)$, since we already know that if they fulfill the above mentioned conditions, they are \textit{fully informative}, whereas if even a single pair is completely missing they are \textit{completely uninformative}.

\paragraph{The subsets with $|B|< n$.} On the other hand, we have demonstrated that the sets with $|B|<n$ are \textit{completely uninformative}, since they necessarily miss at least one full pair.  

\paragraph{The aligned storage register subsets with $|B|=n$.}
The only unauthorized subsets not yet covered by the above cases are the storage register subsets with $|B|=n$, which do not miss an entire signal-noise pair: they need to contain exactly one qubit from each pair, therefore they must have the form
\[
B_{n,p}=\{S_1,S_2,\dots,S_p,N_{p+1},\dots,N_{n-1}, N_n\}
\]
up to permutation symmetry. In other words, they must contain $p\geq 0$ signal qubits, and $q=n-p$ noise qubits, while the alignment is adopted for clarity. 
Since the $B_{n,p}$ are unauthorized, thus cannot be fully informative: we need to find out whether they are partially  or completely uninformative.
Therefore, in the following, we analyze the reduced states of unauthorized subsets of size $|B|=n$ of the storage register ${\mathcal R}_n$. 

\section{Illustrative low-dimensional cases for the storage register}\label{sec:low-dim}

Before turning to the general pattern, it is useful to examine the lowest-dimensional instances explicitly, since these cases make the leakage mechanism transparent.
We start with the case $n=1$: although the protocol is formulated for $n>1$, the exceptional case $n=1$ is instructive because it reveals the basic leakage mechanism.

We present the intermediate steps explicitly, not only for completeness, but also to make the pattern of cancellations and recurrences evident, so that the result can be verified by direct inspection.
\subsection{The case $n=1$: the onset of partial leakage}
For \(n=1\), the density matrix of the encoded state can be written as
\begin{equation}
\rho_{\mathrm{enc}}^{(1)}
=
\ket{\Psi_{\mathrm{enc}}^{(1)}}\!\bra{\Psi_{\mathrm{enc}}^{(1)}}
=
\frac{1}{4}
\sum_{\mu,\nu=0}^3
\alpha_\mu^{-1}\alpha_\nu\,
\Big(\sigma_\mu \ket{\psi}\!\bra{\psi}\sigma_\nu\Big)_A
\otimes
\Big(
\ket{\phi_\mu}\!\bra{\phi_\nu}
\Big)_{S_1N_1}.
\label{eq:rho-enc-n1}
\end{equation}
with
\begin{equation}\label{eq:i1}
\alpha_0=1,
\qquad
\alpha_1=i,
\qquad
\alpha_2=1,
\qquad
\alpha_3=i.
\end{equation}

We know by rule (AUTH1) that the set $\{S_1,N_1\}$ is completely informative. Thus the interesting subsets to be investigated are $\{S_1\}$ and $\{N_1\}$. Hence, we will have to always trace out $A$ and depending on the set, trace out $N_1$ or $S_1$. To this purpose we recall some practical identities. The trace for $A$ yields the scalar
\begin{equation}
\operatorname{Tr}_A\Big(\sigma_\mu \ket{\psi}\!\bra{\psi}\sigma_\nu\Big)=\bra{\psi}\sigma_\nu\sigma_\mu\ket{\psi},
\end{equation}
whereas, tracing out $N_1$ and $S_1$ would yield respectively
\begin{equation}
\operatorname{Tr}_N\!\Big(\ket{\phi_\mu}\!\bra{\phi_\nu}\Big)
=
\frac{1}{2}\sigma_\mu\sigma_\nu,
\qquad
\operatorname{Tr}_S\!\Big(\ket{\phi_\mu}\!\bra{\phi_\nu}\Big)
=
\frac{1}{2}(\sigma_\nu\sigma_\mu)^{\top}.
\label{eq:bell-identities-corrected}
\end{equation}
These yield for the subset $\{S_1\}$
\begin{eqnarray}
\rho_{S_1}^{(1)}
&=&
\frac{1}{4}
\sum_{\mu,\nu=0}^3
\alpha_\mu^{-1}\alpha_\nu\,
Tr_A\Big(\sigma_\mu \ket{\psi}\!\bra{\psi}\sigma_\nu\Big)_A
\otimes
Tr_{N_1}\Big(
\ket{\phi_\mu}\!\bra{\phi_\nu}
\Big)_{S_1N_1}
\\
&=&
\frac{1}{4}
\sum_{\mu,\nu=0}^3
\alpha_\mu^{-1}\alpha_\nu\,
\bra{\psi}\sigma_\nu\sigma_\mu\ket{\psi}\,
\frac{1}{2}(\sigma_\mu\sigma_\nu)
\label{eq:rho-s1-n1}
\end{eqnarray}
and for the subset $\{N_1\}$
\begin{eqnarray}
\rho_{N_1}^{(1)}
&=&
\frac{1}{4}
\sum_{\mu,\nu=0}^3
\alpha_\mu^{-1}\alpha_\nu\,
Tr_A\Big(\sigma_\mu \ket{\psi}\!\bra{\psi}\sigma_\nu\Big)_A
\otimes
Tr_{S_1}\Big(
\ket{\phi_\mu}\!\bra{\phi_\nu}
\Big)_{S_1N_1}
\\
&=&
\frac{1}{4}
\sum_{\mu,\nu=0}^3
\alpha_\mu^{-1}\alpha_\nu\,
\bra{\psi}\sigma_\nu\sigma_\mu\ket{\psi}\,
\frac{1}{2}(\sigma_\nu\sigma_\mu)^{\top}
\label{eq:rho-n1-n1}
\end{eqnarray}
The two expressions differ only by the last factor in the $16$ sum terms.
Since the interference patterns in the entire discussion that follows depend on the signs of the terms and on their cancelation, it is worthwhile to make them explicit. We will now derive 
\begin{itemize}
\item
the table $\textsf C$ of the $16$ scalar coefficients $\alpha_\mu^{-1}\alpha_\nu$, \item
the table $\textsf A$ of the $16$ scalar coefficients $\bra{\psi}\sigma_\nu\sigma_\mu\ket{\psi}$, \item
the tables of the operatorial factors 
\begin{itemize}
\item
$(\sigma_\mu\sigma_\nu)$ (the matrix will be denoted by $\mathsf S$, since we leave the \textit{signal} qubit when tracing  out the noise qubit); 
\item
$(\sigma_\nu\sigma_\mu)^{\top}$ (the matrix will be denoted by ${\mathsf N}$, since we leave the \textit{noise} qubit when tracing out the signal qubit).
\end{itemize}
\end{itemize}
Finally we will perform the pointwise multiplication of those tables:\ the resulting table will contain all the $16$ terms of the sum and we will be able to check for possible cancellations.

\paragraph{The scalar coefficient matrix ${\mathsf C}^{n=1}$.}
The table of the 
$\alpha_\mu^{-1}\alpha_\nu$ coefficients for $n=1$, 
based on (\ref{eq:i1}), is
\begin{equation}\label{eq:alpha-1}
\begin{array}{c|cccc}
\alpha_\mu^{-1}\alpha_\nu & \nu=0 & \nu=1 & \nu=2 & \nu=3\\
\hline
\mu=0 & 1 & i & 1 & i\\
\mu=1 & -i & 1 & -i & 1\\
\mu=2 & 1 & i & 1 & i\\
\mu=3 & -i & 1 & -i & 1
\end{array}
\end{equation}
A decomposition of the table that will turn out to be useful is the following.
\[
\begin{pmatrix}
1 & i & 1 & i\\
-i & 1 & -i & 1\\
1 & i & 1 & i\\
-i & 1 & -i & 1
\end{pmatrix}
=
I_4
+
\begin{pmatrix}
\cdot & i & \cdot & \cdot\\
-i & \cdot & \cdot & \cdot\\
\cdot & \cdot & \cdot & i\\
\cdot & \cdot & -i & \cdot
\end{pmatrix}
+
\begin{pmatrix}
\cdot & \cdot & 1 & \cdot\\
\cdot & \cdot & \cdot & 1\\
1 & \cdot & \cdot & \cdot\\
\cdot & 1 & \cdot & \cdot
\end{pmatrix}
+
\begin{pmatrix}
\cdot & \cdot & \cdot & i\\
\cdot & \cdot & -i & \cdot\\
\cdot & i & \cdot & \cdot\\
-i & \cdot & \cdot & \cdot
\end{pmatrix}.
\]
with $I_4$ the $4\times 4$ identity.
For brevity, we will also use
\[
{\mathsf C}^{(1)} = I_4+{\mathsf C}^{(1)}_1+{\mathsf C}^{(1)}_2+{\mathsf C}^{(1)}_3,
\]
with the matrices defined as above.

\paragraph{The scalar coefficient matrix ${\mathsf A}$.} We recall the rules of the Pauli algebra $XY=iZ$, $YZ=iX$, $ZX=iY$, $\sigma_{\nu}^2=I$ and for $\mu\neq\nu$,
\(
(\sigma_\nu\sigma_\mu) = -(\sigma_\mu\sigma_\nu)
\) with $\mu,\nu\in\{1,2,3\}$.

Hence, explicitly:
\begin{equation}\label{eq:Pauli-algebra}
\begin{array}{cc}
\begin{array}{c|cccc}
\sigma_\mu \sigma_\nu & \nu=0 & \nu=1 & \nu=2 & \nu=3 \\
\hline
\mu=0 & I & X & Y & Z \\
\mu=1 & X & I & XY=iZ & XZ=-iY \\
\mu=2 & Y & YX=-iZ & I & YZ=iX \\
\mu=3 & Z & ZX=iY & ZY=-iX & I
\end{array}
&
\begin{array}{c|cccc}
\sigma_\nu \sigma_\mu & \nu=0 & \nu=1 & \nu=2 & \nu=3 \\
\hline
\mu=0 & I & X & Y & Z \\
\mu=1 & X & I & -iZ & iY \\
\mu=2 & Y & iZ & I & -iX \\
\mu=3 & Z & -iY & iX & I
\end{array}
\end{array}
\end{equation}
Considering the definition of the Bloch vector components
\begin{equation}\label{eq:Bloch-components}
\langle\psi| I|\psi\rangle=1, \quad\langle\psi| X|\psi\rangle=x, \quad\langle\psi| Y|\psi\rangle=y, \quad\langle\psi| Z|\psi\rangle=z.
\end{equation}
the table for the scalar  $\langle\psi|\sigma_\nu\sigma_\mu|\psi\rangle$ can be written as follows.
\begin{equation}\label{eq:scalar}
\begin{array}{c|cccc}
\langle\psi|\sigma_\nu\sigma_\mu|\psi\rangle & \nu=0 & \nu=1 & \nu=2 & \nu=3\\
\hline
\mu=0 & 1 & x & y & z\\
\mu=1 & x & 1 & -iz & iy\\
\mu=2 & y & iz & 1 & -ix\\
\mu=3 & z & -iy & ix & 1
\end{array}
\end{equation}
Writing it as follows
\[
\begin{pmatrix}
1 & x & y & z \\
x & 1 & -iz & iy \\
y & iz & 1 & -ix \\
z & -iy & ix & 1
\end{pmatrix}
= I_4
+
x\,{\mathsf A}_1
+
y\,{\mathsf A}_2
+
z\,{\mathsf A}_3
\]
with
\begin{equation}\label{eq:matrix-A}
{\mathsf A}_1\equiv
\begin{pmatrix}
\cdot & 1 & \cdot & \cdot \\
1 & \cdot & \cdot & \cdot \\
\cdot & \cdot & \cdot & -i \\
\cdot & \cdot & i & \cdot
\end{pmatrix}
\qquad
{\mathsf A}_2\equiv
\begin{pmatrix}
\cdot & \cdot & 1 & \cdot \\
\cdot & \cdot & \cdot & i \\
1 & \cdot & \cdot & \cdot \\
\cdot & -i & \cdot & \cdot
\end{pmatrix}
\qquad
{\mathsf A}_3\equiv
\begin{pmatrix}
\cdot & \cdot & \cdot & 1 \\
\cdot & \cdot & -i & \cdot \\
\cdot & i & \cdot & \cdot \\
1 & \cdot & \cdot & \cdot
\end{pmatrix}.
\end{equation}
highlights the elements associated with individual Bloch components. 

Notice that this table is independent of the multiplicity $n$.
\paragraph{The operator valued matrices.}
The matrix
\(
{\mathsf S} \equiv (\sigma_\mu \sigma_\nu),
\)
whose entries are listed in (\ref{eq:Pauli-algebra}) (left side)\ admits the following decomposition.
\begin{eqnarray}
{\mathsf S} \equiv (\sigma_\mu \sigma_\nu)
&=&
I I_4
+X\,{\mathsf S}_{1}
+Y\,{\mathsf S}_{2}
+Z\,{\mathsf S}_{3}.
\end{eqnarray}
with
\begin{equation}\label{eq:matrix-S}
{\mathsf S}_1\equiv
\begin{pmatrix}
\cdot & 1 & \cdot & \cdot \\
1 & \cdot & \cdot & \cdot \\
\cdot & \cdot & \cdot & i \\
\cdot & \cdot & -i & \cdot
\end{pmatrix}
\qquad {\mathsf S}_2\equiv
\begin{pmatrix}
\cdot & \cdot & 1 & \cdot \\
\cdot & \cdot & \cdot & -i \\
1 & \cdot & \cdot & \cdot \\
\cdot & i & \cdot & \cdot
\end{pmatrix}
\qquad {\mathsf S}_3\equiv
\begin{pmatrix}
\cdot & \cdot & \cdot & 1 \\
\cdot & \cdot & i & \cdot \\
\cdot & -i & \cdot & \cdot \\
1 & \cdot & \cdot & \cdot
\end{pmatrix}.
\end{equation}

The matrix
\(
{\mathsf N} \equiv \big((\sigma_{\nu}\sigma_{\mu})^{\top}\big)_{\mu,\nu=0}^3
\)
obtained transposing the individual elements in the right side table in (\ref{eq:Pauli-algebra}), and recalling that $X^{\top}=X$, $Y^{\top}=-Y$, and  $Z^{\top}=Z$, admits the following decomposition.

\begin{eqnarray}
(\sigma_\nu \sigma_\mu)^{\top}
&=&
I I_4
+X\,{\mathsf N}_{1}
+Y\,{\mathsf N}_{2}
+Z\,{\mathsf N}_{3}.
\end{eqnarray}
with
\begin{equation}\label{eq:matrix-N}
{\mathsf N}_1\equiv
\begin{pmatrix}
\cdot & 1 & \cdot & \cdot \\
1 & \cdot & \cdot & \cdot \\
\cdot & \cdot & \cdot & -i \\
\cdot & \cdot & i & \cdot
\end{pmatrix}
\qquad {\mathsf N}_2\equiv
\begin{pmatrix}
\cdot & \cdot & -1 & \cdot \\
\cdot & \cdot & \cdot & -i \\
-1 & \cdot & \cdot & \cdot \\
\cdot & i & \cdot & \cdot
\end{pmatrix}
\qquad {\mathsf N}_3\equiv
\begin{pmatrix}
\cdot & \cdot & \cdot & 1 \\
\cdot & \cdot & -i & \cdot \\
\cdot & i & \cdot & \cdot \\
1 & \cdot & \cdot & \cdot
\end{pmatrix}.
\end{equation}

\paragraph{Wrapping up.}
We now perform the pointwise multiplication, denoted by $\circ$ and take the sum of the elements of the matrices, defined by
\[
{\mathcal T}({\mathsf M})\equiv \sum_{\mu,\nu=0}^3 {\mathsf M}_{\mu\nu}
\]

and denoting the Bloch components denoted by $b_1=x, b_2=y, b_3=z$) for the set $\{S_1\}$ we get
\[
8\,\rho_{S_1}=\,{\mathcal T}(I_4)\,\sigma_0 + \sum_{j=1}^3 b_j\,{\mathcal T}\Big({\mathsf C}^{(1)}_j\circ {\mathsf A}_j\circ {\mathsf S}_j\Big)\sigma_j
\]
and for the set $\{N_1\}$
\[
8\,\rho_{N_1}=\,{\mathcal T}(I_4)\,\sigma_0 + \sum_{j=1}^3 b_j\,{\mathcal T}\Big({\mathsf C}^{(1)}_j\circ {\mathsf A}_j\circ {\mathsf N}_j\Big)\sigma_j
\]
Now explicitating the result of the pointwise multiplication and the sum we have for $\{S_1\}$
$$
8\,\rho_{S_1}=
I\,{\mathcal T}(I_4)
+x\,X\,{\mathcal T}\!
\left(\begin{array}{cccc}
0 & i & 0 & 0 \\
-i & 0 & 0 & 0 \\
0 & 0 & 0 & i \\
0 & 0 & -i & 0
\end{array}\right)+
yY\,{\mathcal T}\!
\left(\begin{array}{cccc}
0 & 0 & 1 & 0 \\
0 & 0 & 0 & 1 \\
1 & 0 & 0 & 0 \\
0 & 1 & 0 & 0
\end{array}\right)+
zZ\,{\mathcal T}\!
\left(\begin{array}{cccc}
0 & 0 & 0 & i \\
0 & 0 & -i & 0 \\
0 & i & 0 & 0 \\
-i & 0 & 0 & 0
\end{array}\right)
$$
and we see that the first and third matrix elements add up to zero, whereas the second matrix elements add up to 4, hence the expression reduces  to
\[
\rho_{S_1}=\frac{1}{2}\big(I+yY\big).
\]
Thus the unauthorized set $\{S_1\}$ is partially informative.

On the other hand the computation for the subset $\{N_1\}$ yields
\[
8\,\rho_{N_1}=I\,{\mathcal T}(I_4)+
xX\,{\mathcal T}\!\left(\begin{array}{cccc}
0 & i & 0 & 0 \\
-i & 0 & 0 & 0 \\
0 & 0 & 0 & -i \\
0 & 0 & i & 0
\end{array}\right)
+
yY\,{\mathcal T}\!
\left(\begin{array}{cccc}
0 & 0 & -1 & 0 \\
0 & 0 & 0 & 1 \\
-1 & 0 & 0 & 0 \\
0 & 1 & 0 & 0
\end{array}\right)
+
zZ\,{\mathcal T}\!
\left(\begin{array}{cccc}
0 & 0 & 0 & i \\
0 & 0 & i & 0 \\
0 & -i & 0 & 0 \\
-i & 0 & 0 & 0
\end{array}\right)
\]
and we see that the elements of all the three matrices add up to zero, hence
\[
\rho_{N_1}=\frac{1}{2}I.
\]
The unauthorized set $\{N_1\}$ is completely uninformative.

The remarkable finding for $n=1$ is that
although the set $\{S_1\}$ is not authorized and does not allow recovery of the input qubit, it retains a residual dependence on the \(y\)-component of the input Bloch vector. The residual dependence on the input state can be extracted directly by measuring the Pauli-\(Y\) observable on \(S_1\).
This is the first instance of partial leakage.
Yamaguchi and Kempf had already stated that for $n=1$ the clone ${S_1}$ was not completely encrypted, and they established that the protocol should be used for $n>1$. However, as we are going to see later, a similar pattern appears also for higher number of clones when $n$ is odd. First, though, we are going to check the even case $n=2$.

\subsection{The case $n=2$.}

For \(n=2\), the density matrix of the encoded state is

\begin{equation}
\rho_{\mathrm{enc}}^{(2)}
=
\ket{\Psi_{\mathrm{enc}}^{(2)}}\!\bra{\Psi_{\mathrm{enc}}^{(2)}}
=
\frac{1}{4}
\sum_{\mu,\nu=0}^3
\alpha_\mu^{-1}\alpha_\nu\,
\Big(\sigma_\mu \ket{\psi}\!\bra{\psi}\sigma_\nu\Big)_A
\otimes
\Big(
\ket{\phi_\mu}\!\bra{\phi_\nu}
\Big)_{S_1N_1}
\otimes
\Big(
\ket{\phi_\mu}\!\bra{\phi_\nu}
\Big)_{S_2N_2}.
\label{eq:rho-enc-n2}
\end{equation}
with 
\begin{equation}\label{eq:i2}
\alpha_0=1,
\qquad
\alpha_1=\alpha_2=\alpha_3=i.
\end{equation}
Hence
\begin{equation}
\begin{array}{c|cccc}
\alpha_\mu^{-1} \alpha_\nu & \nu=0 & \nu=1 & \nu=2 & \nu=3 \\
\hline
\mu=0 & 1 & i & i & i \\
\mu=1 & -i & 1 & 1 & 1 \\
\mu=2 & -i & 1 & 1 & 1 \\
\mu=3 & -i & 1 & 1 & 1
\end{array}
\end{equation}
and
\[
{\mathsf C}^{(2)} =
\begin{pmatrix}
1 & i & i & i\\
-i & 1 & 1 & 1\\
-i & 1 & 1 & 1\\
-i & 1 & 1 & 1
\end{pmatrix}
=
\begin{pmatrix}
1 & \cdot & \cdot & \cdot\\
\cdot & 1 & \cdot & \cdot\\
\cdot & \cdot & 1 & \cdot\\
\cdot & \cdot & \cdot & 1
\end{pmatrix}
+
\begin{pmatrix}
\cdot & i & \cdot & \cdot\\
-i & \cdot & \cdot & \cdot\\
\cdot & \cdot & \cdot & 1\\
\cdot & \cdot & 1 & \cdot
\end{pmatrix}
+
\begin{pmatrix}
\cdot & \cdot & i & \cdot\\
\cdot & \cdot & \cdot & 1\\
-i & \cdot & \cdot & \cdot\\
\cdot & 1 & \cdot & \cdot
\end{pmatrix}
+
\begin{pmatrix}
\cdot & \cdot & \cdot & i\\
\cdot & \cdot & 1 & \cdot\\
\cdot & 1 & \cdot & \cdot\\
-i & \cdot & \cdot & \cdot
\end{pmatrix}.
\]
or
\[
{\mathsf C}^{(2)} = I_4+{\mathsf C}^{(2)}_1+{\mathsf C}^{(2)}_2+{\mathsf C}^{(2)}_3,
\]

The interesting unauthorized subsets are $\{S_1,N_2\}$ and $\{S_1,S_2\}$ (the behavior of the set $\{S_2,N_1\}$ can be derived from the former by symmetry).

The reduced state for the subset $\{S_1,N_2\}$ is obtained by tracing out $A$ plus one noise qubit from one pair and one signal qubit from the other.
\begin{eqnarray}
\rho_{S_1N_2}
&=&
\frac{1}{4}
\sum_{\mu,\nu=0}^3
\alpha_\mu^{-1}\alpha_\nu\,
Tr_{A}\Big(\sigma_\mu \ket{\psi}\!\bra{\psi}\sigma_\nu\Big)_A
\otimes
Tr_{N_1}
\Big(
\ket{\phi_\mu}\!\bra{\phi_\nu}
\Big)_{S_1N_1}
\otimes
Tr_{S_2}\Big(
\ket{\phi_\mu}\!\bra{\phi_\nu}
\Big)_{S_2N_2}.
\\&=&
\frac{1}{4}
\sum_{\mu,\nu=0}^3
\alpha_\mu^{-1}\alpha_\nu\,
\bra{\psi}\sigma_\nu\sigma_\mu\ket{\psi}\,\frac{1}{2}
(\sigma_\mu\sigma_\nu)\otimes\frac{1}{2}(\sigma_\nu\sigma_\mu)^{\top}.
\end{eqnarray}
The two-signal qubit subset \(\{S_1,S_2\}\) is obtained tracing out $A$ and both the noise qubits $N_1,N_2$.
\begin{eqnarray}
\rho_{S_1S_2}^{(2)}
&=&
\frac{1}{4}
\sum_{\mu,\nu=0}^3
\alpha_\mu^{-1}\alpha_\nu\,
Tr_{A}\Big(\sigma_\mu \ket{\psi}\!\bra{\psi}\sigma_\nu\Big)_A
\otimes
Tr_{N_1}
\Big(
\ket{\phi_\mu}\!\bra{\phi_\nu}
\Big)_{S_1N_1}
\otimes
Tr_{N_2}\Big(
\ket{\phi_\mu}\!\bra{\phi_\nu}
\Big)_{S_2N_2}.
\\&=&
\frac{1}{4}
\sum_{\mu,\nu=0}^3
\alpha_\mu^{-1}\alpha_\nu\,
\bra{\psi}\sigma_\nu\sigma_\mu\ket{\psi}\,\frac{1}{2}
(\sigma_\mu\sigma_\nu)\otimes\frac{1}{2}(\sigma_\mu\sigma_\nu).
\label{eq:rho-s1s2-n2}
\end{eqnarray}
Now, we have already all the necessary factors from the previous section.
For the set $\{S_1,N_{2}\}$ we get
\[
16\,\rho_{S_1N_2}=\,{\mathcal T}(I_4)\,\sigma_0\otimes \sigma_0 + \sum_{j=1}^3 b_j\,{\mathcal T}\Big({\mathsf C}^{(2)}_j\circ {\mathsf A}_j\circ {\mathsf S}_j\circ {\mathsf N}_j\Big)\,\sigma_j\otimes\sigma_j
\]
and for the set $\{S_1,S_{2}\}$ we  have
\[
16\,\rho_{S_1S_2}=\,{\mathcal T}(I_4)\,\sigma_0\otimes \sigma_0 + \sum_{j=1}^3 b_j\,{\mathcal T}\Big({\mathsf C}^{(2)}_j\circ {\mathsf A}_j\circ {\mathsf S}_j\circ {\mathsf S}_j\Big)\,\sigma_j\otimes\sigma_j
\]
It is easy to check that for $j=1,2,3$ the terms ${\mathcal T}\Big({\mathsf C}^{(2)}_j\circ {\mathsf A}_j\circ {\mathsf S}_j\circ {\mathsf N}_j\Big)$ and ${\mathcal T}\Big({\mathsf C}^{(2)}_j\circ {\mathsf A}_j\circ {\mathsf S}_j\circ {\mathsf S}_j\Big)$ are always zero. Hence both the examined reduced states are uninformative
\[
\rho_{S_1N_2}=\frac{1}{4}\big(I\otimes I\big)
\qquad
\rho_{S_1S_2}=\frac{1}{4}\big(I\otimes I\big)
\]
This is the first manifestation of the even-parity cancellation pattern: all Bloch-dependent contributions vanish.
\subsection{The case $n=3$:\ partial leakage}
In the case $n=3$ the interesting unauthorized cases are $\{S_1,N_2,N_3\}$, $\{S_1,S_2,N_3\}$, and $\{S_1,S_2,S_3\}$.

Tracing out the $A$ and the complementary qubits of the storage register, the reduced states can be written.
\begin{eqnarray}
\rho_{S_1N_2N_3}
&=&
\frac{1}{32}
\sum_{\mu,\nu=0}^{3}
\alpha_\mu^{-1}\alpha_\nu\,
\bra{\psi}\sigma_\nu\sigma_\mu\ket{\psi}\,
(\sigma_\mu\sigma_\nu)\otimes
(\sigma_\nu\sigma_\mu)^{\top}\otimes
(\sigma_\nu\sigma_\mu)^{\top}.
\label{eq:rho-s12n3-n3}
\\
\rho_{S_1S_2N_3}
&=&
\frac{1}{32}
\sum_{\mu,\nu=0}^{3}
\alpha_\mu^{-1}\alpha_\nu\,
\bra{\psi}\sigma_\nu\sigma_\mu\ket{\psi}\,
(\sigma_\mu\sigma_\nu)\otimes
(\sigma_\mu\sigma_\nu)\otimes
(\sigma_\nu\sigma_\mu)^{\top}.
\\
\rho_{S_1S_2S_3}
&=&
\frac{1}{32}
\sum_{\mu,\nu=0}^{3}
\alpha_\mu^{-1}\alpha_\nu\,
\bra{\psi}\sigma_\nu\sigma_\mu\ket{\psi}\,
(\sigma_\mu\sigma_\nu)\otimes
(\sigma_\mu\sigma_\nu)\otimes
(\sigma_\mu\sigma_\nu).
\end{eqnarray}
The coefficients $\alpha_{\mu}$ for $n=3$ are
\begin{equation}
\alpha_0=1,
\qquad
\alpha_1=\alpha_3=i,
\qquad
\alpha_2=-1.
\end{equation}
This differs from the \(n=2\) case precisely in the phase of the \(\mu=2\) branch, and this difference will be responsible for the appearance of leakage.
The coefficients $\alpha_\mu^{-1} \alpha_\nu$ are
\begin{equation}
\begin{array}{c|cccc}
\alpha_\mu^{-1} \alpha_\nu & \nu=0 & \nu=1 & \nu=2 & \nu=3 \\
\hline
\mu=0 & 1 & i & -1 & i \\
\mu=1 & -i & 1 & i & 1 \\
\mu=2 & -1 & -i & 1 & -i \\
\mu=3 & -i & 1 & i & 1
\end{array}
\end{equation}
and
\[
{\mathsf C}^{(3)} =
\begin{pmatrix}
1 & i & -1 & i\\
-i & 1 & i & 1\\
-1 & -i & 1 & -i\\
-i & 1 & i & 1
\end{pmatrix}
=
\begin{pmatrix}
1 & \cdot & \cdot & \cdot\\
\cdot & 1 & \cdot & \cdot\\
\cdot & \cdot & 1 & \cdot\\
\cdot & \cdot & \cdot & 1
\end{pmatrix}
+
\begin{pmatrix}
\cdot & i & \cdot & \cdot\\
-i & \cdot & \cdot & \cdot\\
\cdot & \cdot & \cdot & -i\\
\cdot & \cdot & i & \cdot
\end{pmatrix}
+
\begin{pmatrix}
\cdot & \cdot & -1 & \cdot\\
\cdot & \cdot & \cdot & 1\\
-1 & \cdot & \cdot & \cdot\\
\cdot & 1 & \cdot & \cdot
\end{pmatrix}
+
\begin{pmatrix}
\cdot & \cdot & \cdot & i\\
\cdot & \cdot & i & \cdot\\
\cdot & -i & \cdot & \cdot\\
-i & \cdot & \cdot & \cdot
\end{pmatrix}.
\]
or
\[
{\mathsf C}^{(3)} = I_4+{\mathsf C}^{(3)}_1+{\mathsf C}^{(3)}_2+{\mathsf C}^{(3)}_3,
\]

For the sets under study we get
\begin{eqnarray}
32\,\rho_{S_1N_2N_3}&=&\,{\mathcal T}(I_4)\,\sigma_0^{\otimes 3}+ \sum_{j=1}^3 b_j\,{\mathcal T}\Big({\mathsf C}^{(3)}_j\circ {\mathsf A}_j\circ {\mathsf S}_j\circ {\mathsf N}_j\circ {\mathsf N}_j\Big)\,\sigma_j^{\otimes 3}
\\
32\,\rho_{S_1S_2N_3}&=&\,{\mathcal T}(I_4)\,\sigma_0^{\otimes 3} + \sum_{j=1}^3 b_j\,{\mathcal T}\Big({\mathsf C}^{(3)}_j\circ {\mathsf A}_j\circ {\mathsf S}_j\circ {\mathsf S}_j\circ {\mathsf N}_j\Big)\,\sigma_j^{\otimes 3}
\\
32\,\rho_{S_1S_2S_3}&=&\,{\mathcal T}(I_4)\,\sigma_0^{\otimes 3} + \sum_{j=1}^3 b_j\,{\mathcal T}\Big({\mathsf C}^{(3)}_j\circ {\mathsf A}_j\circ {\mathsf S}_j\circ {\mathsf S}_j\circ {\mathsf S}_j\Big)\,\sigma_j^{\otimes 3}
\\
\end{eqnarray}
Developing the pointwise products and summing one finds that the terms ${\mathcal T}(\cdot)$ are always zero except when the number of signal qubits $S_i$ in the set is odd. Thus the reduced states can be written, respectively,
\[
\rho_{S_1N_2N_3}=\frac{1}{8}\big(I^{\otimes 3} - yY^{\otimes 3}\big)
\qquad
\rho_{S_1S_2N_3}=\frac{1}{8}I^{\otimes 3}
\qquad
\rho_{S_1S_2S_3}=\frac{1}{8}\big(I^{\otimes 3}-yY^{\otimes 3}\big)
\]
For $n=3$ the unauthorized sets of $n$ qubits crossing all the pairs, are uninformative if they contain an even number of signals qubits, while they are partially informative if they contain an odd number of signal qubits. The overall sign of the surviving term, negative here, is the $n=3$ instance of the factor $(-1)^{(n-1)/2}$ established in general in Section~\ref{sec:general}.

Taken together, the low multiplicity cases already display the essential structure of the problem. Partial leakage, when present, is highly constrained; it disappears in the first even case and reappears in the odd case only for subsets with odd signal-qubit parity. The next section shows that this pattern persists in full generality.

\section{General classification of subsets of the storage register}\label{sec:general}

The examples of Section~3 suggest that leakage in encrypted cloning is not generic, but highly structured.
We now demonstrate that the parity-dependent patterns observed are valid for generic $n$.

By the reduction established in Section~2, the only unauthorized subsets that remain to be classified are, up to permutation of the pairs, the aligned subsets
\[
B_{n,p}=\{S_1,\dots,S_p,N_{p+1},\dots,N_n\},
\qquad 1\le p\le n.
\]
From now on, we will denote by $p$ the number of signal qubits, by $q$ the number of noise qubits, with $p+q=n$.

We will show that these subsets are completely non-informative unless both $n$ and $p$ are odd, in which case they retain a residual dependence on the input state through the $y$-Bloch component alone.

For these subsets, tracing out the source qubit \(A\) and the complementary qubit in each pair yields the reduced state
\begin{eqnarray}\label{eq:reduced-register}
\rho_{S_1,\dots,S_p,N_{p+1},\dots,N_{p+q}}
&=&
\frac{1}{4}
\sum_{\mu,\nu=0}^3
\alpha_\mu^{-1}\alpha_\nu\,
\bra{\psi}\sigma_\nu\sigma_\mu\ket{\psi}\,\frac{1}{2^{p}}
(\sigma_\mu\sigma_\nu)^{\otimes p}\otimes\frac{1}{2^{q}}\left((\sigma_\nu\sigma_\mu)^{\top}\right)^{\otimes q}.
\end{eqnarray}
This expression makes it possible to separate the contributions associated with the Bloch components

\paragraph{The factor $\alpha_\mu^{-1}\alpha_\nu$.}
From the expressions (\ref{eq:alphas})  for $\alpha_\mu$'s, we have
\[
\alpha_0^{-1}=1,\qquad
\alpha_1^{-1}=\alpha_3^{-1}=-i,\qquad
\alpha_2^{-1}=-i^{-(n+1)}=-(-i)^{n+1}.
\]
Hence, the $n$-dependent matrix
\[
\begin{array}{c|cccc}
\alpha_\mu^{-1} \alpha_\nu & \nu=0 & \nu=1 & \nu=2 & \nu=3 \\
\hline
\mu=0 & 1 & i & -i^{n+1} & i \\
\mu=1 & -i & 1 & i^{n+2} & 1 \\
\mu=2 & -i^{-(n+1)} & -i^{-n} & 1 & -i^{-n} \\
\mu=3 & -i & 1 & i^{n+2} & 1
\end{array}
\]

A useful decomposition of this matrix is the following 
\[
\begin{pmatrix}
1 & i & -i^{\,n+1} & i\\
-i & 1 & i^{\,n+2} & 1\\
-i^{-(n+1)} & -i^{-n} & 1 & -i^{-n}\\
-i & 1 & i^{\,n+2} & 1
\end{pmatrix}
= I_4+{\mathsf C}^{(n)}_1+{\mathsf C}^{(n)}_2+{\mathsf C}^{(n)}_3,
\]
with $I_4$ the identity, and
\[
{\mathsf C}^{(n)}_1
\equiv
\begin{pmatrix}
\cdot & i & \cdot & \cdot\\
-i & \cdot & \cdot & \cdot\\
\cdot & \cdot & \cdot & -i^{-n}\\
\cdot & \cdot & i^{\,n+2} & \cdot
\end{pmatrix}
\quad
{\mathsf C}^{(n)}_2\equiv
\begin{pmatrix}
\cdot & \cdot & -i^{\,n+1} & \cdot\\
\cdot & \cdot & \cdot & 1\\
-i^{-(n+1)} & \cdot & \cdot & \cdot\\
\cdot & 1 & \cdot & \cdot
\end{pmatrix}
\quad
{\mathsf C}^{(n)}_3
\equiv
\begin{pmatrix}
\cdot & \cdot & \cdot & i\\
\cdot & \cdot & i^{\,n+2} & \cdot\\
\cdot & -i^{-n} & \cdot & \cdot\\
-i & \cdot & \cdot & \cdot
\end{pmatrix}.
\]
\paragraph{The factor $\langle\psi|\sigma_\nu\sigma_\mu|\psi\rangle$.}
The constant terms arising from the tracing out of $A$ are independent of $n$.
As already mentioned, the coefficient matrix can be expanded as
$I_4
+
x\,{\mathsf A}_1
+
y\,{\mathsf A}_2
+
z\,{\mathsf A}_3
$,
highlighting the elements associated with individual Bloch components, with the matrices ${\mathsf A}_{j}$ ($j=1,2,3$) given by equation (\ref{eq:matrix-A}).

\paragraph{The factor $(\sigma_\mu \sigma_\nu)^{\otimes p}$.}
Considering the expansion 
\(
(\sigma_\mu \sigma_\nu)
=
I I_4
+X\,{\mathsf S}_{1}
+Y\,{\mathsf S}_{2}
+Z\,{\mathsf S}_{3}.
\)
with the matrices ${\mathsf S}_j$ ($j=1,2,3$) provided by equation (\ref{eq:matrix-S}),
the operator valued matrix $(\sigma_\mu \sigma_\nu)^{\otimes p}$ can be written
\begin{eqnarray}
(\sigma_\mu \sigma_\nu)^{\otimes p}
&=&
I^{\otimes p} I_4
+X^{\otimes p}\,{\mathsf S}_{1}^{\circ p}
+Y^{\otimes p}\,{\mathsf S}_{2}^{\circ p}
+Z^{\otimes p}\,{\mathsf S}_{3}^{\circ p},
\end{eqnarray}
with the repeated pointwise product of a matrix ${\mathsf M}$ defined as ${\mathsf M}_j^{\circ p}\equiv \bigcirc_{\ell=1}^p {\mathsf M}_{j}$. More explicitly
\[
{\mathsf S}_1^{\circ p}\equiv
\begin{pmatrix}
\cdot & 1 & \cdot & \cdot \\
1 & \cdot & \cdot & \cdot \\
\cdot & \cdot & \cdot & i^p \\
\cdot & \cdot & (-i)^p & \cdot
\end{pmatrix}
\qquad {\mathsf S}_2^{\circ p}\equiv
\begin{pmatrix}
\cdot & \cdot & 1 & \cdot \\
\cdot & \cdot & \cdot & (-i)^p \\
1 & \cdot & \cdot & \cdot \\
\cdot & i^p & \cdot & \cdot
\end{pmatrix}
\qquad {\mathsf S}_3^{\circ p}\equiv
\begin{pmatrix}
\cdot & \cdot & \cdot & 1 \\
\cdot & \cdot & i^p & \cdot \\
\cdot & (-i)^p & \cdot & \cdot \\
1 & \cdot & \cdot & \cdot
\end{pmatrix}.
\]

\paragraph{The factor $\left((\sigma_\nu \sigma_\mu)^{\top}\right)^{\otimes q}$.}
Considering the expansion
\(
(\sigma_\nu \sigma_\mu)^{\top}
=
I I_4
+X\,{\mathsf N}_{1}
+Y\,{\mathsf N}_{2}
+Z\,{\mathsf N}_{3}.
\)
with the matrices ${\mathsf N}_j$ ($j=1,2,3$) provided by equation (\ref{eq:matrix-N}),
the operator valued matrix $\left((\sigma_\nu \sigma_\mu)^{\top}\right)^{\otimes q}$ can be written
\begin{eqnarray}
\left((\sigma_\nu \sigma_\mu)^{\top}\right)^{\otimes q}
&=&
I^{\otimes q} I_4
+X^{\otimes q}\,{\mathsf N}_{1}^{\circ q}
+Y^{\otimes q}\,{\mathsf N}_{2}^{\circ q}
+Z^{\otimes q}\,{\mathsf N}_{3}^{\circ q}.
\end{eqnarray}
More explicitly,
\[
{\mathsf N}_1^{\circ q}\equiv
\begin{pmatrix}
\cdot & 1 & \cdot & \cdot \\
1 & \cdot & \cdot & \cdot \\
\cdot & \cdot & \cdot &  (-i)^q  \\
\cdot & \cdot & i^q & \cdot
\end{pmatrix}
\qquad {\mathsf N}_2^{\circ q}\equiv
\begin{pmatrix}
\cdot & \cdot & (-1)^q & \cdot \\
\cdot & \cdot & \cdot & (-i)^q \\
(-1)^q & \cdot & \cdot & \cdot \\
\cdot & i^q & \cdot & \cdot
\end{pmatrix}
\qquad {\mathsf N}_3^{\circ q}\equiv
\begin{pmatrix}
\cdot & \cdot & \cdot & 1 \\
\cdot & \cdot &  (-i)^q & \cdot \\
\cdot & i^q & \cdot & \cdot \\
1 & \cdot & \cdot & \cdot
\end{pmatrix}.
\]
\paragraph{Wrapping up.} Thanks to those conventions, for each $j=1,2,3$, and assuming $p+q=n$ we define
\[
{\mathsf M}_j^{(n,p,n-p)}
\equiv
{\mathsf C}^{(n)}_j
\circ
{\mathsf A}_j
\circ
{\mathsf N}^{\circ (n-p)}_j
\circ
{\mathsf S}^{\circ (p)}_j.
\]
We now take the sum of the elements of the matrices, defined by
\(
{\mathcal T}({\mathsf M})\equiv \sum_{\mu,\nu=0}^3 {\mathsf M}_{\mu\nu}
\).

For \(j=1\), we obtain
\[
{\mathcal T}\!\left({\mathsf M}_1^{(n,p,n-p)}\right)
=
{\mathcal T}
\begin{pmatrix}
\cdot & i & \cdot & \cdot\\
-i & \cdot & \cdot & \cdot\\
\cdot & \cdot & \cdot & i^{\,1-2n+2p}\\
\cdot & \cdot & i^{\,2n-2p-1} & \cdot
\end{pmatrix}
=
i^{\,1-2n+2p}+i^{\,2n-2p-1}=0.
\]
since it has the form  \(i^m+i^{-m}\) with
\(
m=(1-2n+2p),
\)
which is clearly odd. 

For \(j=3\), we obtain
\[
{\mathcal T}\!\left({\mathsf M}_3^{(n,p,n-p)}\right)
=
{\mathcal T}
\begin{pmatrix}
\cdot & \cdot & \cdot & i\\
\cdot & \cdot & i^{\,2p+1} & \cdot\\
\cdot & i^{-2p-1} & \cdot & \cdot\\
-i & \cdot & \cdot & \cdot
\end{pmatrix}=
i^{-2p-1}+i^{\,2p+1}=0.
\]
since this is of the form \(i^m+i^{-m}\), now with
\(
m=2p+1,
\)
which is clearly odd. 

For \(j=2\), we obtain
\[
{\mathsf M}_2^{(n,p,n-p)}
=
\begin{pmatrix}
\cdot & \cdot & -(-1)^{\,n-p} i^{\,n+1} & \cdot\\
\cdot & \cdot & \cdot & i^{\,1-n}\\
-(-1)^{\,n-p} i^{-(n+1)} & \cdot & \cdot & \cdot\\
\cdot & i^{\,n-1} & \cdot & \cdot
\end{pmatrix}.
\]
Accordingly,
\[
{\mathcal T}\!\left({\mathsf M}_2^{(n,p,n-p)}\right)
=
\bigl(1+(-1)^{\,n-p}\bigr)\bigl(i^{\,(n-1)}+i^{\,-(n-1)}\bigr).
\]

The first factor vanishes when \(n-p\) is odd, that is, when \(n\) and \(p\) have opposite parity. If \(n-p\) is even, the first factor equals \(2\). The second factor vanishes when \(n-1\) is odd, i.e. when \(n\) is even, and equals \(2(-1)^{(n-1)/2}\) when \(n\) is odd. Hence the \(y\)-contribution survives if and only if both \(n\) and \(p\) are odd. Overall,
\[
\mathcal{T}\left(\mathrm{M}_2^{(n,p,n-p)}\right)=
\begin{cases}
0, & n \text{ even,} \\
0, & n \text{ odd and } p \text{ even,} \\
4(-1)^{{(n-1)}/{2}}, & n \text{ odd and } p \text{ odd.}
\end{cases}
\]
We therefore obtain the complete reduced-state classification for all aligned unauthorized subsets.
\[
\boxed{
\rho_{S_1,\ldots,S_p,N_{p+1},\ldots,N_n}
=
\frac{1}{2^n}
\begin{cases}
I^{\otimes n},
& n \text{ even or } p \text{ even},\\[1ex]
I^{\otimes n}+(-1)^{{(n-1)}/{2}}\,y\,Y^{\otimes n},
& n \text{ odd and } p \text{ odd}.
\end{cases}
}
\]

\paragraph{Magnitude of the residual dependence.}
The classification above establishes \emph{when} the $y$-component survives, but not how much can be learned from it. Consider the two extreme cases $y=\pm 1$, i.e. an input prepared in an eigenstate of $Y$. The corresponding reduced states of a partially informative subset are
\[
\rho_\pm=\frac{1}{2^n}\Big(I^{\otimes n}\pm(-1)^{(n-1)/2}\,Y^{\otimes n}\Big)
=\frac{1}{2^{\,n-1}}\,\Pi_\pm ,
\]
where $\Pi_\pm$ are the projectors onto the two eigenspaces of $(-1)^{(n-1)/2}Y^{\otimes n}$, each of dimension $2^{n-1}$. The two states are therefore maximally mixed on \emph{orthogonal} supports, and their trace distance equals $1$. An unauthorized subset with $n$ and $p$ both odd consequently discriminates the two $Y$-eigenstates of the input \emph{with certainty}; for a general input, a single measurement of $Y^{\otimes n}$ returns an unbiased estimate of $(-1)^{(n-1)/2}y$. The residual dependence identified above is thus not a marginal effect: along the $y$-axis of the Bloch sphere the leakage is complete.

\paragraph{Locality of the residual dependence.}
The surviving dependence is nevertheless carried entirely by the full-weight correlator $Y^{\otimes n}$. Since this is the only non-identity term of $\rho_B$, tracing out any single qubit of a partially informative subset $B$ removes it, and every proper subset of $B$ is maximally mixed. Extracting the leaked component therefore requires a \emph{joint} measurement on all $n$ qubits of $B$, that is, simultaneous access to one representative of every signal-noise pair. This has a direct architectural consequence: if the components of the storage register are held at independent sites, the residual information is not accessible to any individual site, and its exploitation requires either physically reuniting the $n$ qubits or performing a distributed joint measurement. Confidentiality in encrypted cloning is thus determined not only by which qubits are exposed, but also by whether they can be measured jointly.

\subsection{Summary of the informativeness properties of subsets of the storage register}
In synthesis, the informativeness properties of the subsets of the storage register can be determined as follows.\vspace{5pt}

\begin{proposition}\label{prop:register}
{\bf Parity based informativeness classification of the register subsets.}\\ \vspace{-10pt}
 
Consider the storage register $${\mathcal R}_n=\{S_1,N_1,S_2,N_2,\dots,S_n,N_n\}$$ and a subset $B\subseteq{\mathcal R}_n$, the informativeness properties of $B$ are as follows.
\begin{itemize}
\item If $B$ misses one or more complete pairs, it is completely uninformative.
\begin{itemize}
\item if $|B|<n$, this necessarily happens.
\end{itemize}
\item If $B$ has at least one representative from each signal-noise pair $\{S_i,N_i\}$, i.e. if it has at least size $n$, then
\begin{itemize}
\item when $|B| > n$ the set $B$ is authorized 
\item when $|B|=n$, 
\begin{itemize}
\item if $n$ is even, $B$ is completely uninformative;
\item if $n$ is odd, 
\begin{itemize}
\item if $p$ is even $B$ is completely uninformative;
\item if $p$ is odd $B$ is partially informative;
\end{itemize}
\end{itemize}
\end{itemize}
\end{itemize}
Partial informativeness corresponds to the state $\rho_{B}=\left(I^{\otimes n}+(-1)^{{(n-1)}/{2}}\,y\,Y^{\otimes n}\right)/2^{n}$.
\end{proposition}

\section{Conclusions}\label{sec:conclusions}

We provided a complete classification of the informativeness properties of subsets of the encrypted-clone storage register. Our results show that non-authorized subsets are not all equivalent: while any subset missing a complete clone-noise pair is completely non-informative, aligned subsets of size \(n\) may retain a restricted residual dependence on the input state. This dependence is highly structured: the \(x\)- and \(z\)-Bloch components always cancel, whereas the \(y\)-component survives only when both \(n\) and the number of clone qubits in the subset are odd.

This yields a parity-dependent confidentiality limitation of encrypted cloning. Although the protocol preserves the no-cloning theorem by allowing only a single successful decryption, it does not imply all-or-nothing secrecy across all non-authorized subsets. Rather, confidentiality depends in a subtle way on the interference structure of the encoded state and on the composition of the subset under consideration. 

These findings also suggest, at least at a preliminary level, that the preservation and distribution of encrypted clones may have architectural consequences for storage-oriented implementations, since residual leakage depends on which qubits are jointly exposed. 

Two directions for future work appear especially natural. One is to extend the present classification to subsets that also include the source qubit \(A\), thereby completing the analysis of register informativeness. The other is to examine how the present leakage structure relates to higher-dimensional extensions of encrypted cloning, in particular to the recent generalization proposed by Ceară for arbitrary finite dimensions \cite{ceara2026cloningencryptedquantumstates}. 

Overall, our results show that encrypted cloning should be assessed not only through recoverability, but also through the finer distinction between authorization and residual informativeness. This distinction is essential for understanding encrypted cloning as a quantum storage primitive.



\end{document}